\documentclass[11pt]{elsarticle}

\makeatletter
\long\def\pprintMaketitle{%\clearpage
  \iflongmktitle\if@twocolumn\let\columnwidth=\textwidth\fi\fi
  \resetTitleCounters
  \def\baselinestretch{1}%
  \printFirstPageNotes
  \begin{center}%
 \thispagestyle{pprintTitle}%
 \def\baselinestretch{1}%
    \LARGE\@title\par\vskip18pt
    \normalsize\elsauthors\par\vskip10pt
    \footnotesize\itshape\elsaddress\par\vskip36pt
    \ifvoid\absbox\else\unvbox\absbox\par\vskip10pt\fi
    \ifvoid\keybox\else\unvbox\keybox\par\vskip10pt\fi
    \end{center}%
  \gdef\thefootnote{\arabic{footnote}}%
  }
\makeatother

\makeatletter
\def\ps@pprintTitle{%
  \let\@oddhead\@empty
  \let\@evenhead\@empty
  \let\@oddfoot\@empty
  \let\@evenfoot\@oddfoot
}
\makeatother

\usepackage{graphicx}
\usepackage{subfig}
\usepackage{multirow}
\usepackage[margin=1.25in]{geometry}
\usepackage[usenames,dvipsnames]{color}
\usepackage{url}
\usepackage[colorlinks = true,
            linkcolor = blue,
            urlcolor  = blue,
            citecolor = blue,
            anchorcolor = blue]{hyperref}
\usepackage{lineno}
\modulolinenumbers[10]
\usepackage{siunitx}
\usepackage{comment}
\usepackage{cleveref}
\newenvironment{Abstract}{\begin{quotation} \begin{center}
                       ABSTRACT
     \end{center}\bigskip  }{\end{quotation}}

\def\Acknowledgements{\bigskip  \bigskip \begin{center} \begin{large}
             \bf ACKNOWLEDGEMENTS \end{large}\end{center}}
\newcommand\snowmass{\begin{center}\rule[-0.2in]{\hsize}{0.01in}\\\rule{\hsize}{0.01in}\\
\vskip 0.1in Submitted to the  Proceedings of the US Community Study\\ 
on the Future of Particle Physics (Snowmass 2021)\\ 
Accelerator Frontier Topical Group 05 (AF05)
\rule{\hsize}{0.01in}\\\rule[+0.2in]{\hsize}{0.01in} \end{center}}

\begin{document}
%\pubblock
\snowmass
\begin{frontmatter}
%%%%%%%%%%%%%%%%%%%%%%%%%%%%%%%%%%%%%%%%%%%%%%%%%%%%%%%%%%%%%%%%%%%%
%\title{Expected Beam Backgrounds at Belle~II}
\title{Beam background expectations for Belle~II at SuperKEKB}
%%%%%%%%%%%%%%%%%%%%%%%%%%%%%%%%%%%%%%%%%%%%%%%%%%%%%%%%%%%%%%%%%%%%
\author[instHawaii]{A.~Natochii}\ead{natochii@hawaii.edu}
\author[instHawaii]{T.~E.~Browder}
\author[instDESY]{L.~Cao}
\author[instNagoya]{K.~Kojima}
\author[instWayneState,instKEK]{D.~Liventsev}
\author[instDuke]{F.~Meier}
\author[instKEK,instSOKENDAI]{K.~R.~Nakamura}
\author[instKEK,instSOKENDAI]{H.~Nakayama}\ead{hiroyuki.nakayama@kek.jp}
\author[instDESY]{C.~Niebuhr}
\author[instLjubljanaJSI]{A.~Novosel}
\author[instPisaUNIV,instPisaINFN]{G.~Rizzo}
\author[instRCNP]{S.~Y.~Ryu}
\author[instLjubljanaUniLJ,instLjubljanaJSI]{L.~Santelj}
\author[instKEK]{X.~D.~Shi}
\author[instDESY]{S.~Stefkova}
\author[instUTokyo]{H.~Tanigawa}
\author[instKEK]{N.~Taniguchi}
\author[instHawaii]{S.~E.~Vahsen}\ead{sevahsen@hawaii.edu}
\author[instTriesteUNIV,instTriesteINFN]{L.~Vitale}
\author[instUTokyo]{Z.~Wang}

\address[instHawaii]{University of Hawaii, Honolulu, Hawaii 96822, USA}
\address[instDESY]{Deutsches Elektronen--Synchrotron, 22607 Hamburg, Germany}
\address[instNagoya]{Nagoya University, Nagoya, Aichi 464-8602, Japan}
\address[instWayneState]{Wayne State University, Detroit, Michigan 48202, U.S.A.}
\address[instKEK]{High Energy Accelerator Research Organization (KEK), Tsukuba 305-0801}
\address[instDuke]{Duke University, Durham, North Carolina 27708, U.S.A.}
\address[instSOKENDAI]{SOKENDAI (The Graduate University for Advanced Studies), Hayama 240-0193}
\address[instLjubljanaJSI]{J. Stefan Institute, 1000 Ljubljana, Slovenia}
\address[instPisaUNIV]{Dipartimento di Fisica, Universit\`{a} di Pisa, I-56127 Pisa, Italy}
\address[instPisaINFN]{INFN Sezione di Pisa, I-56127 Pisa, Italy}
\address[instRCNP]{Research Center for Nuclear Physics, Osaka University, Osaka 567-0047, Japan}
\address[instLjubljanaUniLJ]{Faculty of Mathematics and Physics, University of Ljubljana, 1000 Ljubljana, Slovenia}
\address[instUTokyo]{Department of Physics, University of Tokyo, Tokyo 113-0033, Japan}
\address[instTriesteUNIV]{Dipartimento di Fisica, Universit\`{a} di Trieste, I-34127 Trieste, Italy}
\address[instTriesteINFN]{INFN Sezione di Trieste, I-34127 Trieste, Italy}

\end{frontmatter}
\newpage
\begin{Abstract}
\noindent 
The Belle~II experiment at the SuperKEKB electron-positron collider aims to collect an unprecedented data set of $50~{\rm ab}^{-1}$ to study $CP$-violation in the $B$-meson system and to search for Physics beyond the Standard Model (BSM). SuperKEKB is already the world's highest-luminosity collider. In order to collect the planned data set within approximately one decade, the target is to reach a peak luminosity of \SI{6.3e35}{cm^{-2}.s^{-1}} by further increasing the beam currents and reducing the beam-size at the interaction point by squeezing the betatron function down to $\beta^{*}_{\rm y}=\SI{0.3}{mm}$. Beam backgrounds are a key challenge in this context. We estimate the expected background evolution in the next ten years and discuss potential challenges and background mitigation strategies. We find that backgrounds will remain high but acceptable until a luminosity of at least \SI{2.8e35}{cm^{-2}.s^{-1}} is reached at $\beta^{*}_{\rm y}=\SI{0.6}{mm}$. Beyond this luminosity, predictions are highly uncertain, owing to a planned redesign of the interaction region. Improved background estimates with reduced uncertainties for the final, maximum-luminosity operation will require completion of this redesign.
\end{Abstract}

\tableofcontents
%\listoffigures
%\listoftables

\def\thefootnote{\fnsymbol{footnote}}
\setcounter{footnote}{0}
%
%\linenumbers

\section{Executive summary}
\label{sec:ExecutiveSummary}

The Belle~II experiment is a general-purpose particle spectrometer used to study $CP$-violation in the $B$-meson system and to search for Physics beyond the Standard Model (BSM) in decays of $B$-mesons, $D$-mesons and tau leptons as well as in the dark sector. The SuperKEKB collider produces $B$-meson pairs and other particles of interest by colliding electron and positron beams with asymmetric energies at the $\Upsilon(4S)$ resonance.

SuperKEKB is a major upgrade of KEKB, and has been operational since 2016. The machine has already reached a world-record luminosity of \SI{3.8e34}{cm^{-2}.s^{-1}} for a vertical betatron function of $\beta^{*}_{\rm y}=\SI{1.0}{mm}$ at the interaction point (IP), but the goal is to increase the luminosity by another order of magnitude in the coming decade, with a current target peak luminosity of \SI{6.3e35}{cm^{-2}.s^{-1}} for $\beta^{*}_{\rm y}=\SI{0.3}{mm}$. For low-emittance
colliding beams, luminosity is increased both by increasing beam currents and by reducing the beam size at the interaction point, utilizing the so-called nano-beam scheme.

Beam backgrounds are one of the key challenges at SuperKEKB. In the SuperKEKB and Belle~II designs, it was already estimated that several Belle~II subdetectors would be subject to close-to-tolerable backgrounds at peak luminosity. The most vulnerable subdetectors are the Time of Propagation (TOP) particle ID system, and the Central Drift Chamber (CDC). In the TOP, the issue is that high hit rates in the micro-channel-plate photo-multipliers (MCP-PMTs) used to the read out the system's Cherenkov-imaging quartz bars deteriorate with integrated avalanche charge. In the CDC, one key issue is that pattern recognition of charged tracks becomes increasingly difficult as the wire-hit rate increases. 

Given the importance of beam background mitigation to the success of the experiment, we have studied backgrounds extensively in the early phases of SuperKEKB running. Backgrounds from luminosity processes, which are expected to dominate at target luminosity, have been slightly (up to \SI{20}{\%}) lower than expected. Backgrounds from single beams, which are currently dominating, are a factor of a few larger than expected, which is in line with the size of typical machine systematics involved.

To prevent accidental damage, gradual deterioration, or reduced reconstruction performance of Belle~II, background countermeasures such as beam collimation and specialized shielding are used. Single-beam background rates depend very strongly on the collimation system settings. In our previous work on beam backgrounds, the simulation of the SuperKEKB collimation system was substantially improved, and is now deemed reliable~\cite{Natochii2021}.

We report on the current background levels in the different Belle II subdetectors, and on the agreement with simulation. We apply correction factors to the simulation to enforce full agreement with measurements. We then simulate the beam optics of future machine scenarios and predict expected collimator settings. This finally allows us to estimate the expected backgrounds for several future machine scenarios using the same correction factors.

We find that backgrounds should remain high but acceptable until a luminosity of at least \SI{2.8e35}{cm^{-2}.s^{-1}} is reached for $\beta^{*}_{\rm y}=\SI{0.6}{mm}$. The most vulnerable Belle~II detector is the TOP, with a predicted safety factor between 1.4 and 2.9. Beyond this luminosity, predictions are highly uncertain for $\beta^{*}_{\rm y}=\SI{0.3}{mm}$, owing to a potential redesign of the interaction region. Improved background estimates with reduced uncertainties for the final, maximum-luminosity operation will require completion of this redesign.

\section{SuperKEKB and Belle~II}
\label{sec:Belle2Experiment}

In this section, we briefly review the collider and detector sub-systems involved in the beam-induced background analysis. Further detail can be founds in Refs.~\cite{Belle2TDR2010,ADACHI201846,Ohnishi2013}.

\noindent {\it SuperKEKB} is an upgrade of the KEKB accelerator~\cite{nlhep1995,Kurokawa2003,Abe2013}. It is a \SI{3}{km}-circumference asymmetric-energy electron-positron collider with a center-of-mass (CM) energy of $\sqrt{s} = \SI{10.58}{GeV}$ which corresponds to the mass of the $\Upsilon(4S)$ resonance. At the IP, \SI{7}{GeV} electrons stored in the high-energy ring (HER) collide with \SI{4}{GeV} electrons accumulated in the low-energy ring (LER). To reach a collision luminosity of order \SI{1e35}{cm^{-2}.s^{-1}}, SuperKEKB utilizes the so-called nano-beam scheme~\cite{Baszczyk2013}, where the vertical and horizontal beam sizes at the IP are squeezed down to \SI{\sim 50}{nm} and \SI{\sim 10}{\micro m}, respectively with a horizontal crossing angle of \SI{83}{mrad} to avoid the hour-glass effect. The relatively large crossing angle also i) allows for a new final focusing system with superconducting quadrupole magnets (QCS) to reside closer to the IP, ii) allows for separate beam lines for the HER and LER, and iii) avoids combined function IP magnets with large fringe fields. To avoid luminosity degradation caused by beam-beam resonances, dedicated sextupole magnets are used to implement a Crab-Waist collision scheme~\cite{raimondi2007beambeam}, which aligns the vertical waistline of one beam along the trajectory of the other beam at the IP.

The upgrade from KEKB to SuperKEKB included the following major items. We note that the list is not exhaustive.
\begin{itemize}
    \item Short LER dipole magnets were replaced with longer ones.
    \item The interaction region (IR), $\pm\SI{4}{m}$ around the IP, was redesigned. This region hosts the Belle~II detector, the final focusing system and two IR beam pipes.
    \item Beam pipes with a titanium nitride (TiN) coating and antechambers were installed in the LER to reduce the power density of the synchrotron radiation (SR).
    \item A damping ring (DR) was constructed, to reduce the positron beam emittance.
    \item The radio-frequency (RF) system was modified to enable higher beam currents.
\end{itemize}

\begin{figure}[htbp]
\begin{center}
\includegraphics[width=\linewidth]{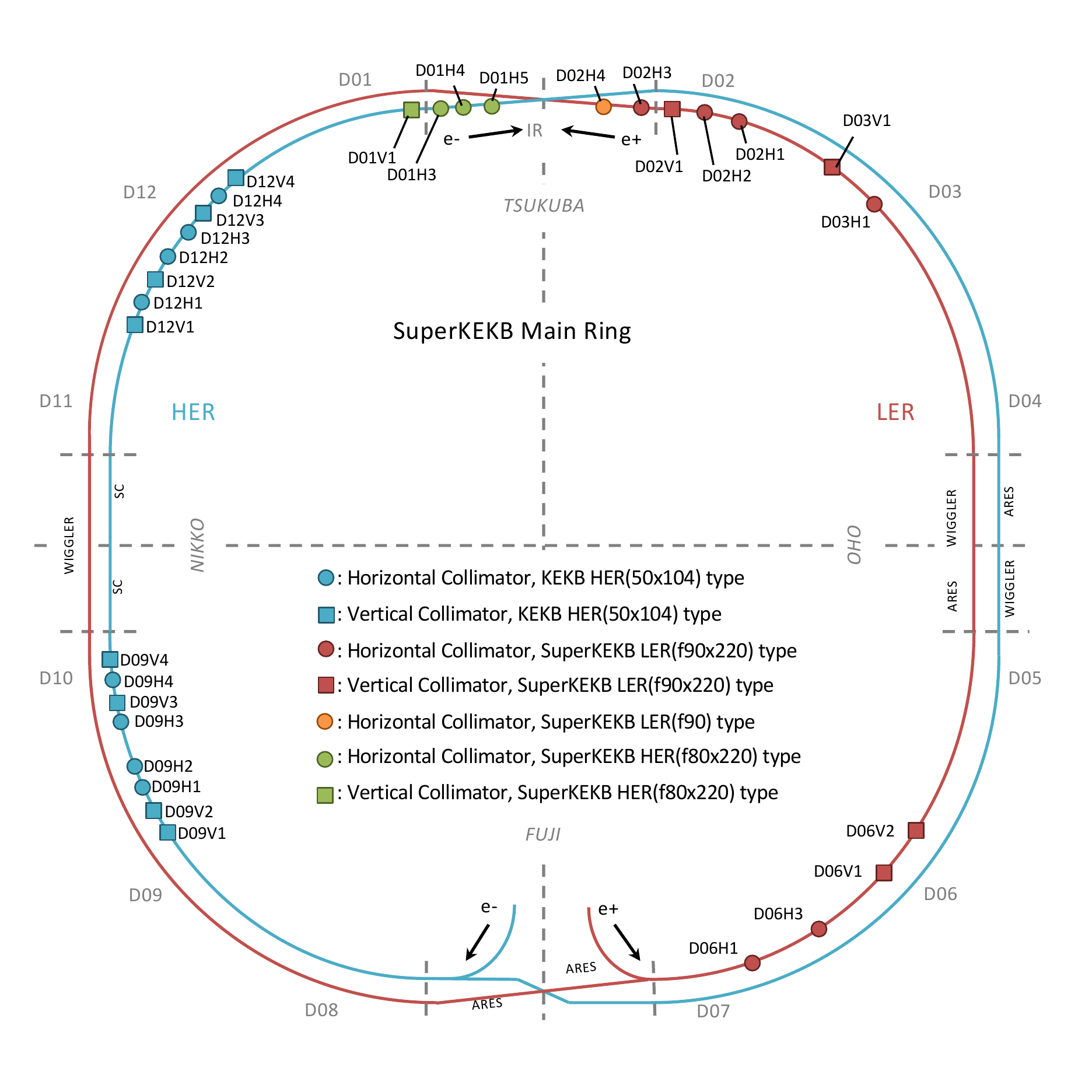}
\end{center}
\caption{\label{fig:fig1}SuperKEKB collimation system as of 2021. The letters V and H in the collimator names indicate vertical and horizontal collimators, respectively. Each ring is divided into twelve sections, referred to as D01 through D12.}

\end{figure}

To avoid radiation damage to the Belle~II detector, which surrounds the IR, and to avoid QCS quenches due to stray beam particles, a set of movable beam collimators is installed around each ring. There are 11 collimators in the LER and 20 in the HER, with current locations as shown in Figure~\ref{fig:fig1}. There are two main types of collimators with different geometries: KEKB collimators are asymmetric and have only one jaw, while SuperKEKB collimators are symmetric with jaws on both sides. More details about the collimators' geometry and material composition can be found in Ref.~\cite{Ishibashi2020,Natochii2021}.

The vacuum system of the collider is designed to effectively mitigate i) higher order mode (HOM) power losses, ii) heat and gas loads due to the large SR power and photon density, iii) the electron cloud and fast ion effects in the LER and HER, respectively. A distributed pumping system based on multilayer non-evaporable getter (NEG) strips~\cite{SUETSUGU2008153} is used to keep the vacuum pressure at the level of \SI{1e-7}{Pa}, which is required to achieve an hours-long beam-gas lifetime. To monitor the vacuum level in the beam pipe, cold cathode gauges (CCGs) are placed approximately every \SI{10}{m} along each ring.

There are two major upgrades of the machine planned in the next ten years: the Long Shutdown~1 (LS1) in 2023, and the Long Shutdown~2 (LS2) around 2027, which is under discussions among collaborators. Our possible future upgrade of the detector is strongly linked to upgrades of the machine. The most crucial upgrades to be implemented are discussed later in the text.

\noindent {\it The Belle~II detector}~\cite{ADACHI201846} is a general purpose particle spectrometer, optimized for precise measurements of $B$-meson pairs via their decay products. The detector must maintain the Belle level of performance~\cite{Abashian2002,Brodzicka2012}, even with the reduced center of mass boost, while operating in a much higher-background environment, which tends to reduce detector performance and longevity. Belle~II replaced a number of Belle sub-systems to satisfy this requirement, and has better vertexing and particle identification performance than Belle. Belle~II consists of several nested sub-detectors surrounding the interaction point (IP) of the two colliding beams, which is within a \SI{1}{\cm} radius beryllium beam pipe. The Belle~II subdetector closest to the IP is the two-layer pixel detector (PXD). All 16 modules in the first layer (L1) of the PXD have been installed, but the second layer (L2) is installed only partially, with 4 of 24 planned modules. During the LS1, a full PXD will be installed. The PXD is surrounded by four layers (L3-6) of the double-sided strip silicon vertex detector (SVD). Both PXD and SVD are surrounded by the central drift chamber (CDC) filled with a $\rm He(\SI{50}{\%}) + C_{2}H_{6}(\SI{50}{\%})$ gas mixture. The CDC consists of 56~layers with 14336~sense wires of either axial or stereo orientation for precise measurements of charged particle trajectories. The charged-particle identification system is based on two sub-detectors: a Time-of-Propagation detector (TOP) and an Aerogel Ring Imaging Cherenkov counter (ARICH) in the barrel and forward endcap regions of Belle~II, respectively. The TOP is composed of \SI{2}{\cm} thick quartz bars with conventional and atomic layer deposition (ALD) types of microchannel plate photomultipliers (MCP-PMTs), which are arranged in 16 readout slots. The ARICH consists of \SI{4}{\cm} thick focusing aerogel radiators and Hybrid Avalanche Photo Detectors (HAPDs), which are grouped in 18 segments. For precise energy and timing measurements of particles an electromagnetic calorimeter (ECL) is installed in the barrel and both endcaps. It is composed of 8736 CsI(Tl) crystals and located inside a superconducting solenoid that provides a \SI{1.5}{T} magnetic field. Outside the coil, a $K^{\rm 0}_{\rm L}$ and muon detector (KLM) is installed. The KLM has 12 and 14 scintillator strip layers read out by silicon PMTs in the forward (FWD) and backward (BWD) endcaps, respectively. The two innermost KLM barrel layers also utilize scintillators, while the 13 outermost barrel layers consist of glass-electrode resistive plate chambers (RPCs).

\noindent {\it A dedicated background monitoring system}
was used in the first commissioning phase of SuperKEKB in 2016~\cite{Lewis2019}. Various detectors measured dose rates around the IR beam pipe and in the accelerator tunnel outside Belle~II. Much of the system was retired when Belle~II was installed. Dedicated background detectors currently in use include:

\begin{itemize}
    \item Diamond sensor-based detectors (Diamonds) monitor the radiation dose rate around the IR beam pipe, see Fig.~\ref{fig:fig3}. Four sensors are a part of the fast beam abort system~\cite{Bacher2021}. 
    \item The sCintillation Light And Waveform Sensors (CLAWS) detector system, based on plastic scintillators and silicon photomultipliers, is primarily used for monitoring Belle~II backgrounds in time with beam injection into the main ring~\cite{Gabriel2021}, and has recently been included into the beam abort logic. Due to its excellent timing performance, CLAWS can trigger a beam abort $\sim\SI{10}{\micro s}$ earlier than Diamonds, on average.
    \item Compact Time Projection Chambers (TPCs) provide directional measurements of the fast neutron flux in the accelerator tunnel~\cite{schueler2021application}.
    \item $\rm He^{3}$ tubes are used for the thermal neutron counting around Belle~II~\cite{deJong2017}. 
\end{itemize}

\section{Beam backgrounds and countermeasures}
\label{sec:BeamBackgroundsAndCountermeasures}

Here, we briefly review the main sources of beam-induced backgrounds, usually simply referred to as beam backgrounds for short. Beam backgrounds can arise from particle losses from the beam during machine operation, leading to electromagnetic (EM) showers. Secondary particles from such showers then cause higher hit rates and other detrimental effects in Belle~II. While this is the typical example, any aspect of machine operation, even the intentional collision of beams, which result in higher Belle~II occupancy or radiation dose, can be considered a beam background. Since SuperKEKB is designed to operate at much higher beam currents and collision luminosities than its predecessor KEKB, the IR backgrounds are also expected to be much higher.

\paragraph{Particle scattering} 
Beam particles circulating in SuperKEKB may undergo single-beam processes caused by i) scattering with residual gas molecules due to Bremsstrahlung and Coulomb interactions, and ii) Coulomb scattering between particles in the same beam bunch, commonly referred to as the Touschek effect. These processes lead to the scattered particles falling out of the stable orbit and reaching the physical or dynamic aperture of the machine. This in turn creates EM showers when these stray particles hit the beam pipe walls. The beam-gas background rate is proportional to the vacuum pressure and the beam current, while the Touschek background is proportional to the beam current squared and inversely proportional to the number of bunches and beam size.

\paragraph{Colliding beams}
While particle collisions at the IP are the goal of SuperKEKB, there are several undesirable collision processes that have very high cross sections but are of little interest for physics measurements. Examples are radiative Bhabha scattering ($e^+ e^- \rightarrow e^+ e^- \gamma$) and two-photon processes ($e^+ e^- \rightarrow e^+ e^- e^+ e^-$). These processes increase the Belle~II occupancy and radiation dose, and we refer to these increases as luminosity background. The rate of this background component is proportional to the luminosity.

\paragraph{Synchrotron radiation (SR)}
X-rays emitted when electrons or positrons pass through the strong magnetic field of the QCS or dipole magnets upstream and downstream of the IP. SR can potentially damage the inner layers of the vertex detector. Because SR power is proportional to the beam energy squared and magnetic field strength squared, X-rays from the HER are the main concern.

\paragraph{Beam injection}
Since the beam lifetime, which is limited mainly by Touschek losses, is of order \SIrange{10}{60}{\min}, SuperKEKB is operated using a so-called top-up injection scheme where the bunches are continuously refilled to keep the current stable. Due to optics mismatches between the injection chain and the main ring, injected particle bunches do not enter the aperture of the machine perfectly, resulting in higher particles losses from injected bunches until they stabilize ($\sim\SI{10}{ms}$).

\paragraph{Catastrophic beam losses}
During otherwise stable machine operation unexplained beam instabilities and beam losses may occasionally occur in one of the rings. Such losses can occur in one shot at a specific location around the ring due to injection kicker errors or due the beam interacting with dust particles in the beam pipe. In extreme cases, high intensity EM showers can lead to detector or collimators damage or superconducting magnet quenches. Usually only a few such catastrophic beam loss events happen per year.

\paragraph{Head-tail instabilities}
In linear accelerators (linacs), when the dense bunch of charged particles travels through the accelerator structure, the front part of the bunch (head) induces an EM wakefield which deflects the rear part of the bunch (tail). After some time the tail starts to oscillate around the head trajectory, which remains unperturbed. These oscillations may cause particle losses on the aperture of the machine. This process is referred to as Beam Break-Up (BBU) instability. In a synchrotron machine like SuperKEKB, the RF field induces a longitudinal focusing of the bunch which creates a collective transverse motion with head-tail modes of betatron and synchrotron oscillation coupling. Below a certain bunch current this oscillation remains stable while at the threshold value the growth rate of the tail is faster than the synchrotron period so that the head-tail modes are destroyed and the beam becomes unstable. This process is referred to as the Transverse Mode Coupling Instability (TMCI)~\cite{Gareyte2001}. Since the TMCI threshold depends on the most narrow and steep aperture in the ring, the bunch current threshold is usually calculated considering only collimators. It can be defined as follows~\cite{Gareyte2001,Chao1999}:
\begin{equation}
I_{\rm th} = \frac{C_{\rm 1}f_{\rm s}E/e}{\sum_{\rm j}\beta_{\rm j}k_{\rm j}},
    \label{eq:eq1}
\end{equation}
where $I_{\rm th}$ is the upper limit on the bunch current, $f_{\rm s} = \SI{2.13}{kHz}$ or $f_{\rm s} = \SI{2.80}{kHz}$ is the synchrotron frequency for the LER or HER, respectively, $E$ is the beam energy, $e$ is the unit charge, $\beta_{\rm j}$ and $k_{\rm j} \sim d^{-\frac{3}{2}}$ are the beta function and kick factor, as a function of the aperture $d$~\cite{Yue2016physdesign}, for the $j$th collimator, respectively. Usually, the constant $C_{\rm 1}$ is equal to 8 or $2\pi$~\cite{Chao1999}, however, for SuperKEKB, $C_{\rm 1} \approx 4\pi$ is assumed to be a more realistic value based on recent machine studies~\cite{OhmiIshibashi2022}. Additional machine elements with a variable aperture, such as the IR beam pipe, can be included into Equation~\ref{eq:eq1} by extending the sum over $\beta_{\rm j}k_{\rm j}$ to include terms both for collimators and these additional elements.

 In the last decade, we have developed a comprehensive set of countermeasures against each of these background sources. We use collimators at specific locations to suppressed single-beam and injection backgrounds by stopping stray particles before they reach the IR. Vacuum scrubbing helps us to reduce the vacuum pressure in each ring, thus lowering the beam-gas scattering rate~\cite{Shibata2019}. A heavy metal shield was installed to protect the inner detectors against EM showers from beam losses in the IR. A polyethylene+lead shield inside the ECL~\cite{Beaulieu2019} protects the ECL crystals and photodiodes from neutrons. Similarly, the inner layers of the ARICH were replaced with a boron-doped polyethylene shield against neutrons.  The incoming beam pipes of the IR collimate most of SR photons and suppress their reflections thanks to the design geometry and a ridge structure on the inner beam pipe surface~\cite{Ohnishi2013}. To protect the vertex detector against residual SR, the inner surface of the IP beryllium beam pipe is coated with a \SI{10}{\micro m} thick gold layer.  The newly constructed DR reduces the emittance of injected positrons, thereby suppressing injection beam losses. We apply an injection veto on the Belle~II Level-1 trigger, to avoid recording high beam losses $\sim\SI{10}{ms}$ after each injection. In addition, we continuously perform injection chain tuning to improve the injection efficiency and to reduce injection background. To protect Belle~II against abnormal beam losses, our Diamonds-based abort system has been complemented by two new systems based on CLAWS detectors near the IR and CsI-crystals around each ring. The CLAWS system provides an additional fast beam abort signal, while the CsI-based sensors help to locate the initial beam loss locations along the ring.
 
\section{Belle~II background status}
\label{sec:Belle2BackgroundStatus}

The Belle~II background level measured in June 2021 is shown in Figure~\ref{fig:fig4}. The data was recorded with a betatron function at the IP of $\beta^{\rm *}_{\rm y}=\SI{1}{mm}$, beam currents of $I_{\rm LER} = \SI{732.6}{mA}$ and $I_{\rm HER} = \SI{647.2}{mA}$, with ring-averaged beam pipe pressures of $\overline P_{\rm LER} = \SI{88}{nPa}$ and $\overline P_{\rm HER} = \SI{24}{nPa}$, $n_{\rm b} = 1174$ bunches in each ring, and a collision luminosity of $\mathcal{L} = \SI{2.6e34}{cm^{-2}s^{-1}}$. Current background rates in Belle~II are acceptable and in most cases well below the limits listed in Table~\ref{tab:tab1}. TOP is the most vulnerable detector. 
%Rates in KLM appear high, but this is an analysis artifact: the barrel KLM RPCs have a high pedestal rate associated with the noise floor. A dedicated  analysis that suppresses the pedestal in KLM-RPC layers is in progress. The KLM is a rather robust detector, with gradual, temporary efficiency reduction but no permanent damage at high rates.

%%%%%%%%%%%%%%%%%%%%%%%%%%%%%%%%%%%%%%%%%%%%%%%%%%%%%%%%%%%%%%%%%%%%%%%%%
\begin{table}[htbp]
\centering
    \caption{\label{tab:tab1}Background rate limits for different Belle~II detector sub-systems. The third column shows the total measured background rate in June~2021 excluding the pedestal rate. TOP limits before/after LS1 are related to the replacement of TOP conventional PMTs planned for the LS1. The upper background rate limit quoted for the Diamond read-out electronics can be increased by selecting a lower signal amplification. The KLM detector limit corresponds to the muon reconstruction efficiency drop of about \SI{10}{\%}.}
    \begin{tabular}{lcccc}
    \hline\hline
    Detector & \multicolumn{3}{c}{BG rate limit} & Measured BG\\
    \hline
    Diamonds & \multicolumn{3}{c}{\SIrange{1}{2}{rad/s}} & $<\SI{125}{mrad/s}$\\
    PXD & \multicolumn{3}{c}{\SI{3}{\%}} & \SI{0.11}{\%}\\
    SVD L3, L4, L5, L6 & \multicolumn{3}{c}{\SI{4.7}{\%}, \SI{2.4}{\%}, \SI{1.8}{\%}, \SI{1.2}{\%}} & $<\SI{0.22}{\%}$\\
    CDC & \multicolumn{3}{c}{\SI{200}{kHz/wire}} & \SI{27}{kHz/wire}\\
    ARICH & \multicolumn{3}{c}{\SI{10}{MHz/HAPD}} & \SI{0.5}{MHz/HAPD}\\
    Barrel KLM L3 & \multicolumn{3}{c}{\SI{50}{MHz}} & \SI{3.8}{MHz}\\
    & \multicolumn{2}{c}{non-luminosity BG} & \multicolumn{1}{c}{luminosity BG}&\\\cline{2-4}
    & before LS1 & after LS1 & per $10^{35}\rm~cm^{-2}s^{-1}$&\\\cline{2-4}
    TOP ALD & \SI{3}{MHz/PMT} & \SI{5}{MHz/PMT} & \SI{0.9}{MHz/PMT} & \SI{2}{MHz/PMT}\\
    \hline\hline
    \end{tabular}
\end{table}
%%%%%%%%%%%%%%%%%%%%%%%%%%%%%%%%%%%%%%%%%%%%%%%%%%%%%%%%%%%%%%%%%%%%%%%%%

Figure~\ref{fig:fig4} shows that dominant backgrounds for all sub-systems are the LER single-beam (beam-gas and Touschek) and luminosity backgrounds. The HER single-beam background is at the level of \SI{10}{\%} for most of the detectors except for the ARICH which is placed in the FWD side of the Belle~II hence sees more forward-directed particle losses in the IR from the electron beam.

The current level of the SR is of no concern in terms of occupancy for the innermost layers of the vertex detector. However, in the case of a large increase, SR may cause inhomogenities in PXD module irradiation, which would make it more difficult to compensate by adjusting the operation voltages of the affected modules.

Although currently at an acceptable level, some detectors started to see single-event upsets (SEU) of FPGAs electronics boards in 2021. These SEUs are presumably from neutrons created in EM showers.

Studies during the same running period revealed beam instabilities that cause vertical emittance blow up in the LER. This occured at lower bunch currents than predicted by standard TMCI theory, limiting the bunch current at the time. Machine studies showed that the instability threshold, see Eq.~\ref{eq:eq1}, not only depends on the collimator settings, but also on the vertical beam size inside the QCS beam pipe structure. This observation implies that the beam instability observed in the LER is not simply a TMCI phenomenon, and might force us to eventually open collimators at higher beam currents and accept higher backgrounds. The betatron function in the QCS depends on the vertical beam size at the IP, and will increases by an order of magnitude if beams at the IP are squeezed maximally in the future, exacerbating this effect, and making it critical to fully understand an mitigate.

\section{Beam background simulation}
\label{sec:MonteCarloSimulation}

We use several dedicated Monte-Carlo (MC) simulation tools to predict beam backgrounds at different machine settings, and to investigate possible beam background mitigation measures.

For the simulation of beam-gas and Touschek scattering, we fist use the Strategic Accelerator Design (SAD)~\cite{SAD2022} software developed at KEK. Bunches of scattered particles ($e^{\rm +}$ or $e^{\rm -}$) are generated at different locations around the ring and tracked for 1000 machine turns. The code collects the coordinates of particles, which reach the aperture of the machine, i.e. the inner beam pipe wall or collimators. We recently improved the simulation of how beam particles interact with collimators.  We have implemented a more realistic collimator profile and now also include particle scattering off of collimator materials. The measured vacuum pressure distribution around the ring is now also used to improve the beam-gas background simulation. Further details about the multi-turn particle tracking in SAD for SuperKEKB can be found in Ref.~\cite{Natochii2021}. 

Next, particles lost from the beam in SAD are passed to Geant4~\cite{Agostinelli2003,Allison2006,Allison2016} for further simulation. We only keep particles lost within the interaction region, out to \SI{30}{m} from the IP. The Geant4 software is embedded in the Belle~II Analysis Software Framework (basf2)~\cite{Kuhr2019}. A realistic description of the geometry of the detector and accelerator tunnel is used to simulate the resulting electromagnetic showers and secondary particles produced inside and outside the detector. Energy deposits in detectors are digitized, and can then be compared against dedicated single-beam and luminosity background measurements.

To produce the MC samples for i) the luminosity background due to radiative Bhabha scattering and two-photon process, and ii) the SR background, we use the same Geant4-based software and geometry, but SAD is not required.

The current accuracy of our beam background simulation is illustrated by Figure~\ref{fig:fig5}, which shows ratios of measured to simulated background rates, so-called data/MC ratios. The ratios are shown separately for different background components. These ratios are now within one order of magnitude of unity, which is a substantial improvement compared to the early SuperKEKB commissioning phases in 2016~\cite{Lewis2019} and 2018~\cite{liptak2021measurements}. 

Since the neutron shielding around Belle~II is not ideal, neutrons can penetrate the detector and cause performance degradation. Our MC simulation predicts fast neutrons traveling from the accelerator tunnel towards Belle~II due to i) single-beam losses at the collimators closest to the IR ($\sim\SI{15}{m}$ from the IP), and ii) radiative Bhabha scattering producing photons parallel to the two beams at the IP. These photons follow the beams out of Belle~II, resulting in localized hot-spots in the beam pipes just outside of the detector ($\sim\SI{10}{m}$ from the IP)~\cite{schueler2021application}.

\section{Background predictions}
\label{sec:BackgroundPrediction}

In this section, we predict the Belle~II background levels expected in the next decade. Our estimates are based on the SuperKEKB Roadmap-2020 program developed by the accelerator team. The machine is expected to gradually increase the luminosity, mainly by squeezing the beam size at the IP and by increasing the beam currents. We evaluate three machine configurations expected in the next years, with predicted machine parameters listed in Table~\ref{tab:tab2}:

\begin{itemize}
    \item \textbf{Setup-1}: an intermediate configuration of the machine, between the currently used optics and conditions expected after LS1;
    \item \textbf{Setup-2}: machine configuration after LS1 but before LS2;
    \item \textbf{Setup-3}: target beam parameters at the design machine lattice to reach an integrated luminosity of about \SI{50}{ab^{-1}}.
\end{itemize}

%%%%%%%%%%%%%%%%%%%%%%%%%%%%%%%%%%%%%%%%%%%%%%%%%%%%%%%%%%%%%%%%%%%%%%%%%
\begin{table}[htbp]
\centering
    \caption{\label{tab:tab2}Predicted SuperKEKB parameters, expected to be achieved by the specified date. $\beta^{\rm *}$, $\mathcal{L}$, $I$, $BD_{\rm int}$, $\overline P$, $n_b$, $\varepsilon$, $\sigma_{\rm z}$ and $CW$ stand for the betatron function at the interaction point, luminosity, beam current, integrated beam dose, average beam pipe gas pressure, number of bunches, equilibrium beam emittance, bunch length and Crab-Waist sextupoles, respectively.}
    \begin{tabular}{lcccccc}
    \hline\hline
    Parameter & Setup-1 & Setup-2 & Setup-3\\
    \hline
    Date & Jan 2023 & Jan 2027 & Jan 2031 \\
    $\beta^{\rm *}_{\rm y}$(LER/HER) [mm] & 0.8/0.8 & 0.6/0.6 & 0.27/0.3 \\
    $\beta^{\rm *}_{\rm x}$(LER/HER) [mm] & 60/60 & 60/60 & 32/25 \\
    $\mathcal{L}$~[$\times 10^{35}\rm~cm^{-2}s^{-1}$] & 1.0 & 2.8 & 6.3 \\
    $I$(LER/HER) [A] & 1.66/1.20 & 2.52/1.82 & 2.80/2.00 \\
    $BD_{\rm int}$~[kAh] & 10 & 45 & 93 \\
    $\overline P$(LER/HER) [nPa] & 93/23 & 48/17 & 33/15\\
    $n_b$ [bunches] & 1370 & 1576 & 1761 \\
    $\varepsilon_{\rm x}$(LER/HER) [nm] & 4.5/4.5 & 4.6/4.5 & 3.3/4.6 \\
    $\varepsilon_{\rm y}/\varepsilon_{\rm x}$(LER/HER) [\%] & 1/1 & 1/1 & 0.27/0.28 \\
    $\sigma_{\rm z}$(LER/HER) [mm] & 7.58/7.22 & 8.27/7.60 & 8.25/7.58 \\
    $CW$ & ON & OFF & OFF \\
    \hline\hline
    \end{tabular}
\end{table}
%%%%%%%%%%%%%%%%%%%%%%%%%%%%%%%%%%%%%%%%%%%%%%%%%%%%%%%%%%%%%%%%%%%%%%%%%

We also consider a variation of Setup-2, which we call \textbf{Setup-2*} with $CW$~=~ON. This machine configuration may result in reduced beam backgrounds and raise the LER TMCI bunch current current limit (which normally results from narrow vertical collimator setting that are required to to keep beam backgrounds under control). Setup-2* is based on the proposal by SuperKEKB experts~\cite{Oide2021} to install a so-called non-linear collimator (NLC)~\cite{Faus2006} in the OHO-section of the LER, see Fig.~\ref{fig:fig1}. In this zero-dispersion region, about \SIrange{700}{800}{m} from the IP, a pair of skew sextupole magnets could be installed, $\sim\SI{0.5}{[2\pi]}$ apart in the vertical betatron phase advance, to create an angular kick in the vertical plane, $\Delta p_{\rm y} \sim (y^{2}-x^{2})$, mainly to halo particles. Deflected particles can then be stopped by a regular vertical collimator placed in between the two nonlinear elements at $\sim\SI{0.25}{[2\pi]}$ phase advance from the sextupoles. The NLC would be opened much wider than typical vertical collimators in order to clean the beam halo efficiently. Since the betatron function of this collimator is small $\sim\SI{3}{m}$, the wake field contribution of the NLC to the machine impedance and hence to the TMCI threshold is negligible, see Eq.~\ref{eq:eq1}. Thus, we can tighten this collimator and relax other vertical collimators to increase the TMCI bunch-current limits while keeping IR beam backgrounds at acceptable levels. The angular kick induced by the sextupoles also affects beam halo particles in both planes, therefore the NLC stops stray particles in both planes, significantly reducing IR beam losses. The simulation of Setup-2* is based on Setup-2 but has an LER lattice compatible with the nonlinear collimator.

The collimation system for each setup listed in Table~\ref{tab:tab2} has been carefully optimized in simulation based on the following considerations: wide open collimators increase beam losses in the IR, while too narrow collimators reduce the beam lifetime and strongly perturb the beam through TMCI. Therefore, each collimator should be set at an aperture that optimally balances backgrounds, lifetimes, and instabilities. More details regarding the collimation system optimization procedure can be found in Ref.~\cite{Natochii2021}. Here, we use Equation~\ref{eq:eq1} to calculate the TMCI bunch current threshold, and include the effect of the QCS wake field, which reduces the bunch current threshold, see Sec.~\ref{sec:Belle2BackgroundStatus}.

Figures~\ref{fig:fig6},~\ref{fig:fig7}~and~\ref{fig:fig8} show simulated Belle~II beam-induced backgrounds  for Setup-1, Setup-2, and Setup-2*, respectively. Rates have been scaled by the data/MC ratios shown in Figure~\ref{fig:fig5} to correct for systematic underestimates of the measured rates observed in past measurements. The predicted background levels for all setups are below the limits listed in Table~\ref{tab:tab1}. Comparing the results for Setup-2 and Setup-2*, we see that the NLC works as expected in our simulation, and reduces IR backgrounds while raising the bunch current limit. At the target machine parameters, the luminosity background dominates, while the LER beam-gas background is kept low due to expected vacuum scrubbing reducing the beam pipe pressure. 
%While the ARICH detector has projected backgrounds that may exceed its limit, we suspect this is a temporary analysis artefact, which is also partially indicated by the abnormally high data/MC ratios for this detector, see Figs.~\ref{fig:fig5},~\ref{fig:fig6}~and~\ref{fig:fig7}.
%While the ARICH and KLM detectors have projected backgrounds that may exceed their limits, but we suspect these are temporary analysis artifacts (for ARICH) and noise floor masking issues (for the RPCs in the KLM), which is also partially indicated by the abnormally high data/MC ratios for these detectors, see Figs.~\ref{fig:fig5},~\ref{fig:fig6}~and~\ref{fig:fig7}. 
While the ARICH limit for the neutron fluence is $\sim \SI{1e12}{neutrons/cm^{2}}$, our Monte-Carlo simulation shows that the expected flux is $\sim \SI{2.5e10}{neutrons/cm^{2}/year}$ which would allows safe detector operation for more than 10~years.

Figure~\ref{fig:fig10} illustrates the predicted evolution of the background rate in the TOP ALD-type MCP-PMTs, starting from current machine settings through Setup-3. The predicted backgrounds in the TOP, which is the most vulnerable detector, are well below the limits for most of the setups. Here, for the collimation system optimization, we first calculate the TMCI bunch current threshold taking into account collimator wake fields only for all setups. We then re-calculate the bunch current thresholds, including also the QCS wake field terms into the sum in Equation~\ref{eq:eq1}. For Setup-1 and Setup-2, the TMCI requirement remains satisfied even with the QCS wake field included. For Setup-3 that is not the case. For Setup-3, only a single vertical LER collimator (D02V1) is used. Therefore, the TMCI bunch current threshold based on collimators and QCS wake fields forces us to open D02V1 much wider than the IR aperture, resulting in TOP rates that exceed the limit in Setup-3. The more realistic TMCI threshold consideration increases LER beam-gas and LER Touschek backgrounds by factors of 100 and 50, respectively. Since we do not have the final version of the machine lattice for the target luminosity, and because of large uncertainties due to beam instability constraints, the total TOP background for Setup-3 is shown in Figure~\ref{fig:fig10} with large systematic uncertainties. The lower and upper edges of the uncertainty band are the TOP rates obtained for the optimized and opened (due to TMCI) D02V1 aperture, respectively. The total background for Setup-3
%The data point itself is simply the arithmetic mean of the predicted rates in those two extreme scenarios, and 
represents a highly uncertain estimate of the Belle~II background at target beam parameters based on the design machine lattice, which %cannot be used anymore and 
should be revisited.

\section{Discussion}
\label{sec:Discussions}

The most recent SuperKEKB luminosity record of \SI{3.81e34}{cm^{-2}.s^{-1}} was reached on December~23, 2021~\cite{KEK2021} at a factor of 6.7 lower product of beam currents for a luminosity 3.2 times higher than at the PEP-II collider~\cite{PEPII2006,Wienands2007}. Currently, SuperKEKB and Belle~II are operating stably. Beam-induced background rates are well below detector limits and do not yet prevent us from increasing beam currents further. This opens a unique opportunity for Belle~II researchers to study rare physics processes at the world highest data collection rate. Although the backgrounds are currently under control, and the forecast presented here looks promising, several other difficulties can also limit beam currents and hence luminosity.

In this section, we i) list open questions and difficulties related to stable machine operation with acceptable background levels, ii) review ongoing activities for further background mitigation, and iii) discuss options for improving the background predicted for the final target beam parameters of SuperKEKB at $\mathcal{L} = \SI{6.3e35}{cm^{-2}.s^{-1}}$. Below each issue is described in terms of its positive (+) or negative (--) expected impact on beam stability, IR backgrounds and/or collision luminosity.

\begin{itemize}
    \item Beam instabilities
    \begin{itemize}
        \item[(--)] Our preliminary estimates show that at the target beam parameters it may be challenging to safely run the experiment due to the low TMCI bunch current threshold at the narrow collimator apertures. Thus, we might be forced to open some collimators, which will increase the IR background exceeding the detector limits.
        \item [(+)] By increasing the gain of the bunch-by-bunch feedback system, by increasing the chromaticity of vertical tune, and/or by the beam-beam effect during collisions the beam instabilities due TMCI effect could be partially cured.
    \end{itemize}
    A set of dedicated machine studies is planned to check this hypothesis experimentally by squeezing the beam at the IP further and measuring the real instability threshold, and how it depends on the aforementioned mitigation measures. A preliminary analysis of the first impedance measurements in 2022 suggests that the QCS wake field effect may be smaller than we simulated. The rate uncertainty shown in Figure~\ref{fig:fig10} for Setup-3 should therefore conservatively cover the range of possible future outcomes.
    \item Non-linear collimation
    \begin{itemize}
        \item [(+)] Our analysis shows that the use of a vertical collimator between a pair of skew sextupole magnets could be beneficial for suppressing the IR background and raising the bunch current threshold.
        \item [(--)] However, the NLC installation would require a rearrangement of the machine components before the wigglers in the OHO section of the LER, which would require a year without beam operation.
    \end{itemize}
    This scenario is under investigations and discussion.
    \item Design lattice
    \begin{itemize}
        \item [(--)] At the design machine lattice without the Crab-Waist scheme used in Setup-3, the target beam currents will be difficult or even impossible to reach because of the short beam lifetime ($<$\SI{10}{min})~\cite{Morita2015}. 
        \item [(+)] Resolving the dynamic aperture and specific luminosity degradation due to the implementation of the Crab-Waist scheme into the SuperKEKB design lattice will allow us to operate the machine stably at the target beam currents and luminosity.
    \end{itemize}
    
    These issues have triggered many studies, investigations and discussions about a possible machine lattice redesign during the LS2. Alternative scenarios to a major redesign of the IR with a very long stop for LS2, are also being evaluated, balancing all pros and cons, such as the option of longer-term machine operation at Setup-2 (Setup-2*) or at intermediate settings between Setup-2 and Setup-3.
    
    %These issues have triggered many studies, investigations and discussions about a possible machine lattice redesign during the LS2. Alternative scenarios to a major re-design of the IR with a very long stop for LS2, are also being evaluated, balancing all pros and cons, like the option of longer-term machine operation at Setup-2 (Setup-2*) or at intermediate settings between Setup-2 and Setup-3.
    %We suggest the Belle~II experiment to also consider the option of longer-term machine operation at Setup-2 (Setup-2*) or at intermediate settings between Setup-2 and Setup-3, 
    %It is motivated by several reasons: backgrounds are expected to remain acceptable, at the first look the machine lattice does not have issues to be implemented in terms of collision performances and beam instabilities, and the luminosity might be at the level of \SI{3e35}{cm^{-2}.s^{-1}}, which is only half of the target value. This scenario would still allow us to collect a large physics data sample by continuously running the machine, avoiding a long delay due to the LS2.
    
    \item Detector shielding
    \begin{itemize}
        \item [(+)] We plan to install additional shielding around Belle~II to suppress the flux of neutrons originating from the accelerator tunnel. These neutrons cause aging of ECL photodiodes and other detector components, and lead to disruptive SEU events. 
        \item [(+)] Additional IR bellows shielding around the QCS is already under construction, and may reduce single-beam and luminosity backgrounds by up to $\SI{50}{\%}$.
        \item [(+)] A new IP beam pipe with an additional gold layer and slightly modified geometry to reduce the amount of the backscattered SR is planned to be installed during the LS1.
    \end{itemize}
    
    \item Backgrounds and detector limits
    \begin{itemize}
        \item [(--)] As we listed in Table~\ref{tab:tab1}, the background limit for the CDC is at the level of \SI{200}{kHz/wire}. However, for the CDC trigger the limit is estimated as \SI{150}{kHz/wire}. Thus, if the CDC background exceeds this rate in the future, the trigger should be upgraded.
        \item [(--)] In our background prediction analysis, we assume that the vacuum scrubbing will improve the residual gas pressure during the machine operation as expected. However, if we unexpectedly reach a saturation value of the vacuum pressure for Setup-2 or Setup-3 that is higher than the value considered in our simulation, then the predicted beam-gas background will increase.
        \item [(+)] During the LS1 we plan to replace the TOP short lifetime conventional MCP-PMTs due to their quantum efficiency degradation. This upgrade will let us operate the detector at a single-beam background level up to \SI{5}{MHz/PMT}.
        \item [(--)] The ECL detector is robust against backgrounds, and we do not have a hard background rate limit. However, its energy resolution slowly degrades as background rates increase. A dedicated ECL analysis is still in development.
        \item [(--)] Bad injection performance may require wider collimator apertures and hence increase Belle~II backgrounds. Therefore, we continuously tune the injection chain of the KEK accelerator complex to improve the injection and mitigate the related beam losses. So far, we have some margin for injection backgrounds on top of the total storage background due to single-beam and luminosity beam losses.
    \end{itemize}
    \item Collimators
    \begin{itemize}
        \item [(+)] An accurate collimator aperture control, at the level of less than \SI{50}{\micro m}, leading to positioning stability, and reproducibility is needed for optimal background mitigation and for correct simulation, which is important for a comprehensive understanding of the machine. A new position read out system for D02V1 was recently installed to improve the collimator control.
        \item [(+)] Hybrid collimator heads made of different material layers may be more robust against abnormal beam hits. To study this, a new low-impedance, hybrid collimator was installed in the LER and will be tested in 2022.
    \end{itemize}
\end{itemize}

\section{Conclusions}
\label{sec:Conclusions}

We have reviewed SuperKEKB and Belle~II upgrades intended to control backgrounds and allow stable operation at a world record collision luminosity of about \SI{4e34}{cm^{-2}.s^{-1}} for $\beta^{*}_{\rm y}=\SI{1.0}{mm}$. Further machine and detector improvements are foreseen in order to reach the target beam parameters. In the next decade, beam-induced backgrounds in Belle~II are expected to remain acceptable until at least $\beta^{*}_{\rm y}=\SI{0.6}{mm}$. This statement assumes the baseline plan of replacing the short-lifetime conventional MCP-PMTs in the TOP detector. Installing additional shielding during the two long shutdowns in 2023 and around 2027 could reduce backgrounds further. There are several important uncertainties that could affect our background forecast in either direction, and which will require further studies and refinement. Due to the issues discussed above related to the design machine lattice and beam instabilities, it is too early to make accurate predictions, but backgrounds could exceed the limit at target beam parameters in Setup-3 for $\beta^{*}_{\rm y}=\SI{0.3}{mm}$. Thus, several machine operation schemes, instability and background countermeasures, and upgrades of the experiment are under consideration in order to collect an integrated luminosity of the order of \SI{50}{ab^{-1}} by the 2030s. We are closely collaborating with EU, US and Asian accelerator laboratories on optimizing upgrades of SuperKEKB and reaching the target luminosity.

\Acknowledgements

We thank the SuperKEKB accelerator and optics groups for the excellent machine operation and for sharing their lattice files; the KEKB vacuum group for the collimation system maintenance;  the KEK computing team for on-site support; our Belle and Belle~II colleagues for the detector operation; K.~Oide, T.~Koga, Y.~Suetsugu, T.~Ishibashi and D.~Zhou (KEK) for their fresh ideas, assistance and constructive discussions regarding the non-linear collimation, triggering scheme, future machine settings and beam instabilities. This work was supported by the U.S. Department of Energy (DOE) via Award Number DE-SC0007852 and via U.S. Belle~II Operations administered by Brookhaven National Laboratory (DE-SC0012704).

\bibliographystyle{JHEP}
\bibliography{myreferences}

\newpage
%%%%%%%%%%%%%%%%%%%%%%%%%%%%%%%%%%%%%%%%%%%%%%%%%%%%%%%%%%%%%%%%%%%%%%%%%%%
%%%%%%%%%%%%%%%%%%%%%%%%%%%%%%%%%%%%%%%%%%%%%%%%%%%%%%%%%%%%%%%%%%%%%%%%%%%
%%%%%%%%%%%%%%%%%%%%%%%%%%%%%%%%%%%%%%%%%%%%%%%%%%%%%%%%%%%%%%%%%%%%%%%%%%%
\begin{figure}[htbp]
\begin{center}
\includegraphics[width=\linewidth]{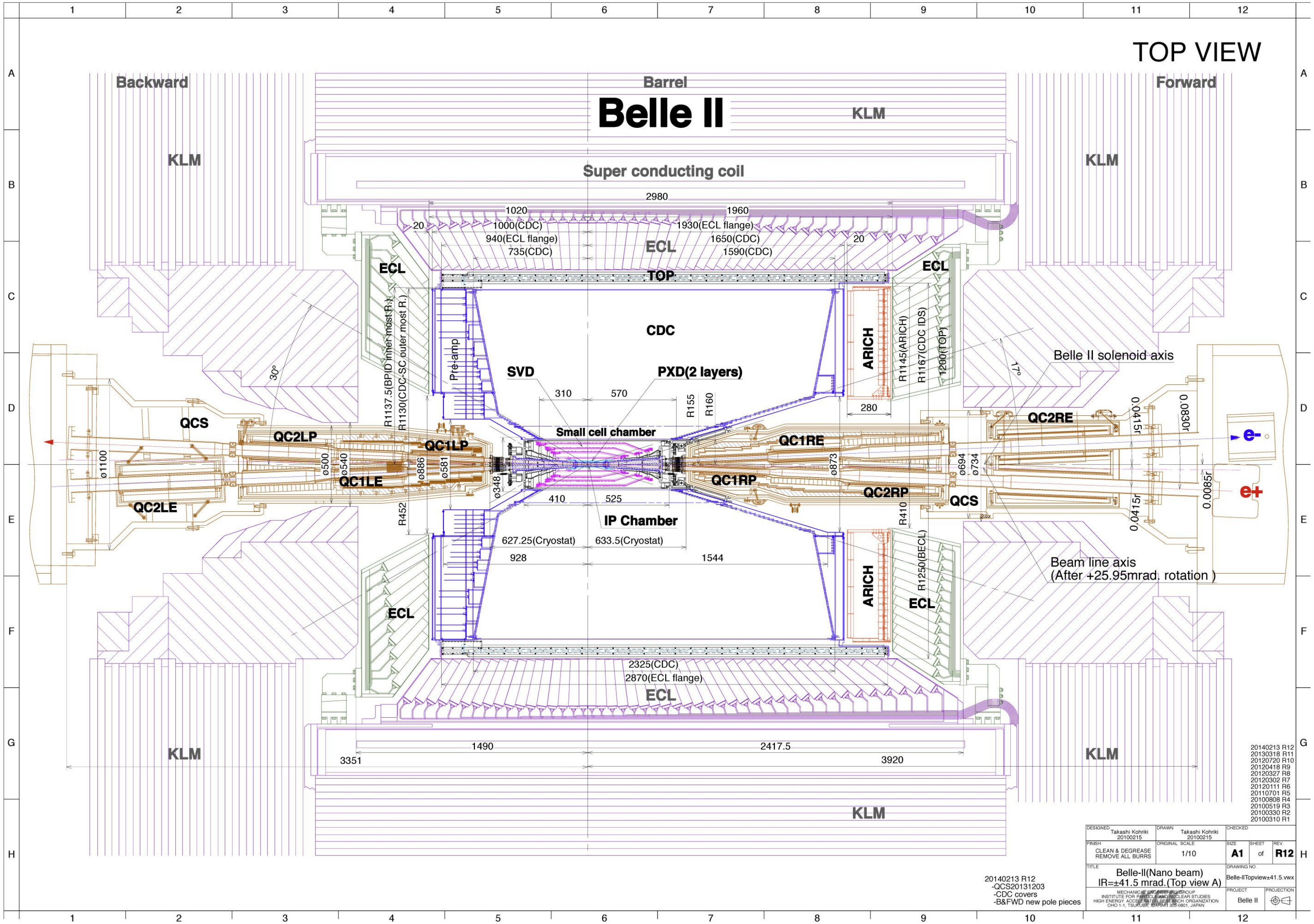}
\end{center}
\caption{\label{fig:fig2}Top view of the Belle~II detector~\cite{ADACHI201846}.}

\end{figure}
%%%%%%%%%%%%%%%%%%%%%%%%%%%%%%%%%%%%%%%%%%%%%%%%%%%%%%%%%%%%%%%%%%%%%%%%%%%
%%%%%%%%%%%%%%%%%%%%%%%%%%%%%%%%%%%%%%%%%%%%%%%%%%%%%%%%%%%%%%%%%%%%%%%%%%%
\begin{figure}[htbp]
\begin{center}
\includegraphics[width=\linewidth]{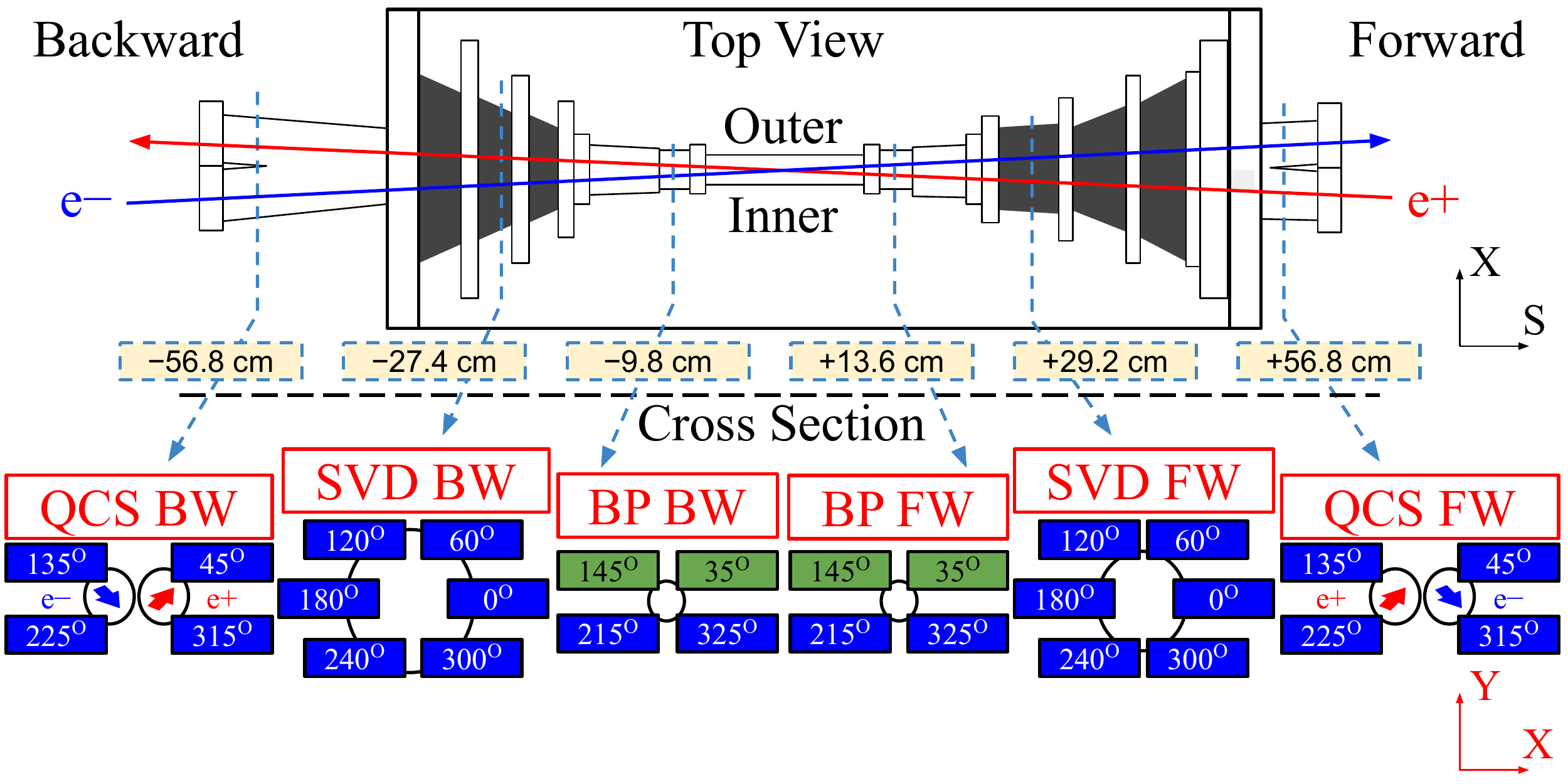}
\end{center}
\caption{\label{fig:fig3}Diamond detector configuration in the IR. The detectors' azimuth angles are indicated in rectangles. The Diamonds highlighted in blue and green are used for dose rate monitoring at a \SI{10}{Hz} readout rate and reserved for the fast beam abort at \SI{400}{kHz}, respectively. The distance from the IP along the beam axis for each group of Diamonds is shown in dashed rectangles.}

\end{figure}
%%%%%%%%%%%%%%%%%%%%%%%%%%%%%%%%%%%%%%%%%%%%%%%%%%%%%%%%%%%%%%%%%%%%%%%%%%%
%%%%%%%%%%%%%%%%%%%%%%%%%%%%%%%%%%%%%%%%%%%%%%%%%%%%%%%%%%%%%%%%%%%%%%%%%%%
\begin{figure}[htbp]
\begin{center}
\includegraphics[width=\linewidth]{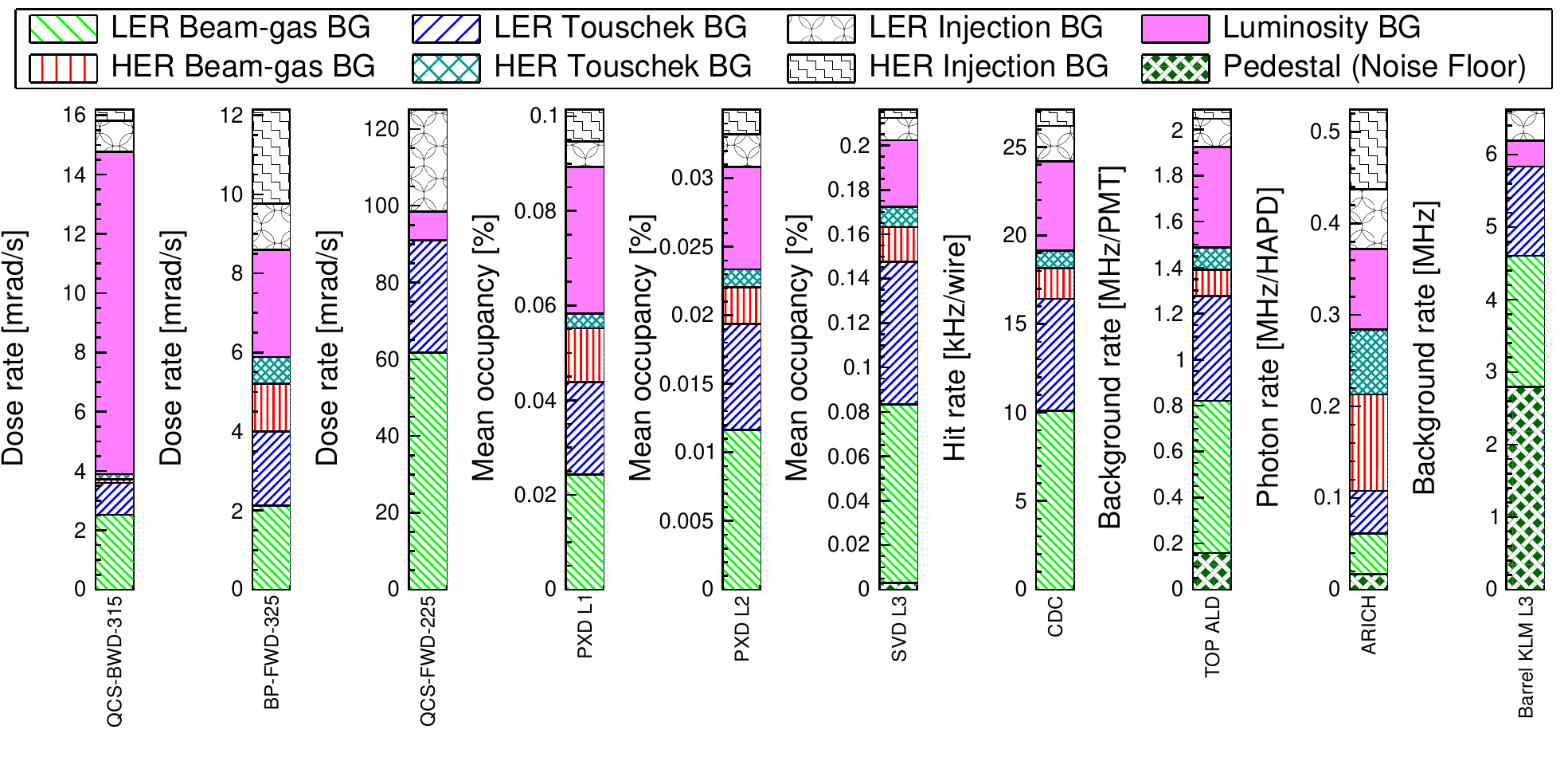}
\end{center}
\caption{\label{fig:fig4}Measured Belle~II background in June~2021. Each column is a stacked histogram. The PXD SR background is too low ($< \SI{1e-3}{\%}$) to be displayed. QCS-BWD-315, BP-FWD-325 and QCS-FWD-225 indicates backward QCS, beam pipe and forward QCS Diamond detectors, respectively, with the higher dose rate. Barrel KLM L3 corresponds to the innermost RPC layer in the barrel region of the KLM detector. TOP ALD shows the averaged background over ALD-type MCP-PMTs, slots from 3 to 8.}

\end{figure}
%%%%%%%%%%%%%%%%%%%%%%%%%%%%%%%%%%%%%%%%%%%%%%%%%%%%%%%%%%%%%%%%%%%%%%%%%%%
%%%%%%%%%%%%%%%%%%%%%%%%%%%%%%%%%%%%%%%%%%%%%%%%%%%%%%%%%%%%%%%%%%%%%%%%%%%
\begin{figure}[htbp]
\centering
\subfloat[\label{secD:fig5a}LER single-beam background.]{\includegraphics[width=0.5\linewidth]{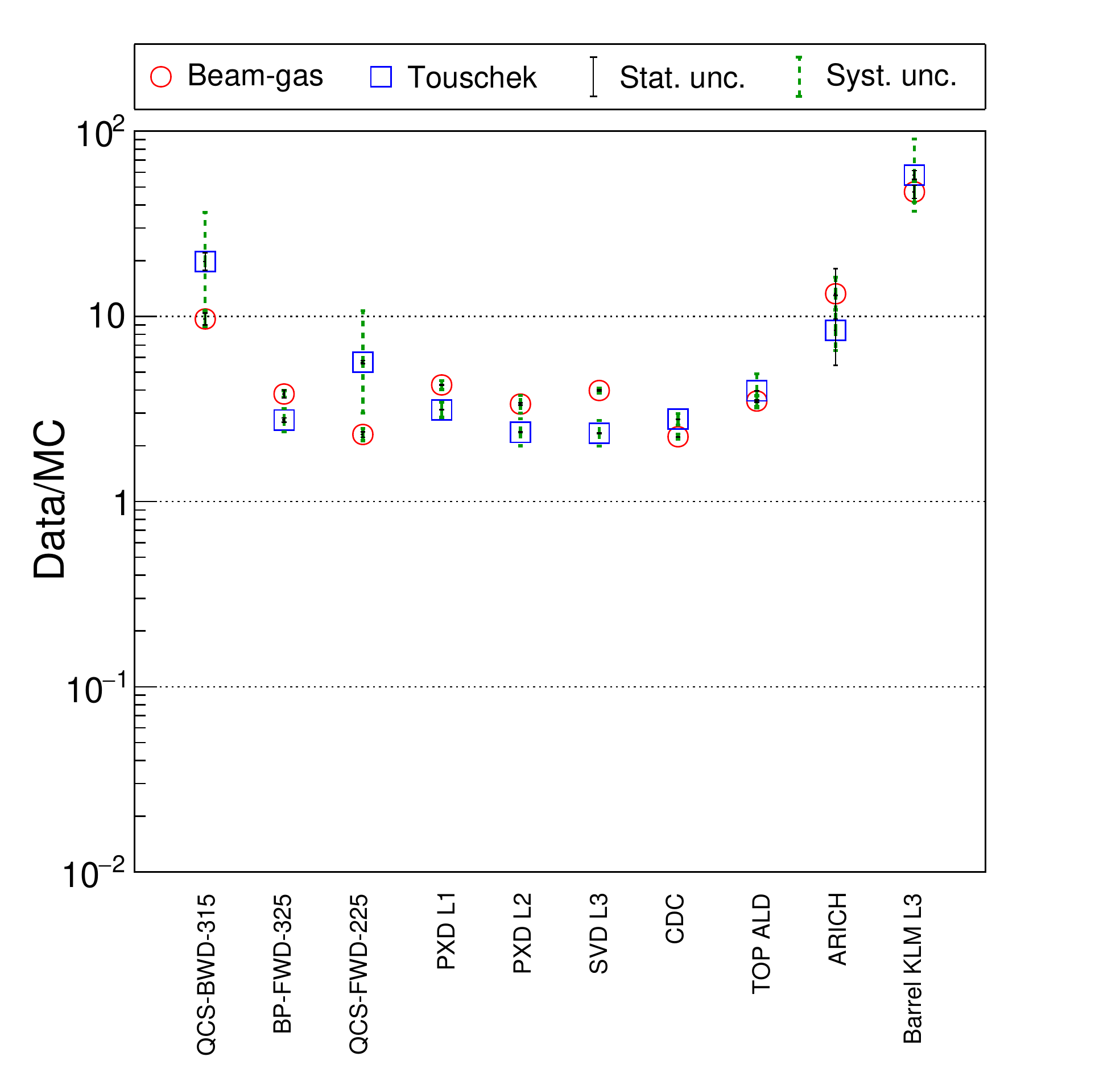}}
\hfill
\subfloat[\label{secD:fig5b}HER single-beam background.]{\includegraphics[width=0.5\linewidth]{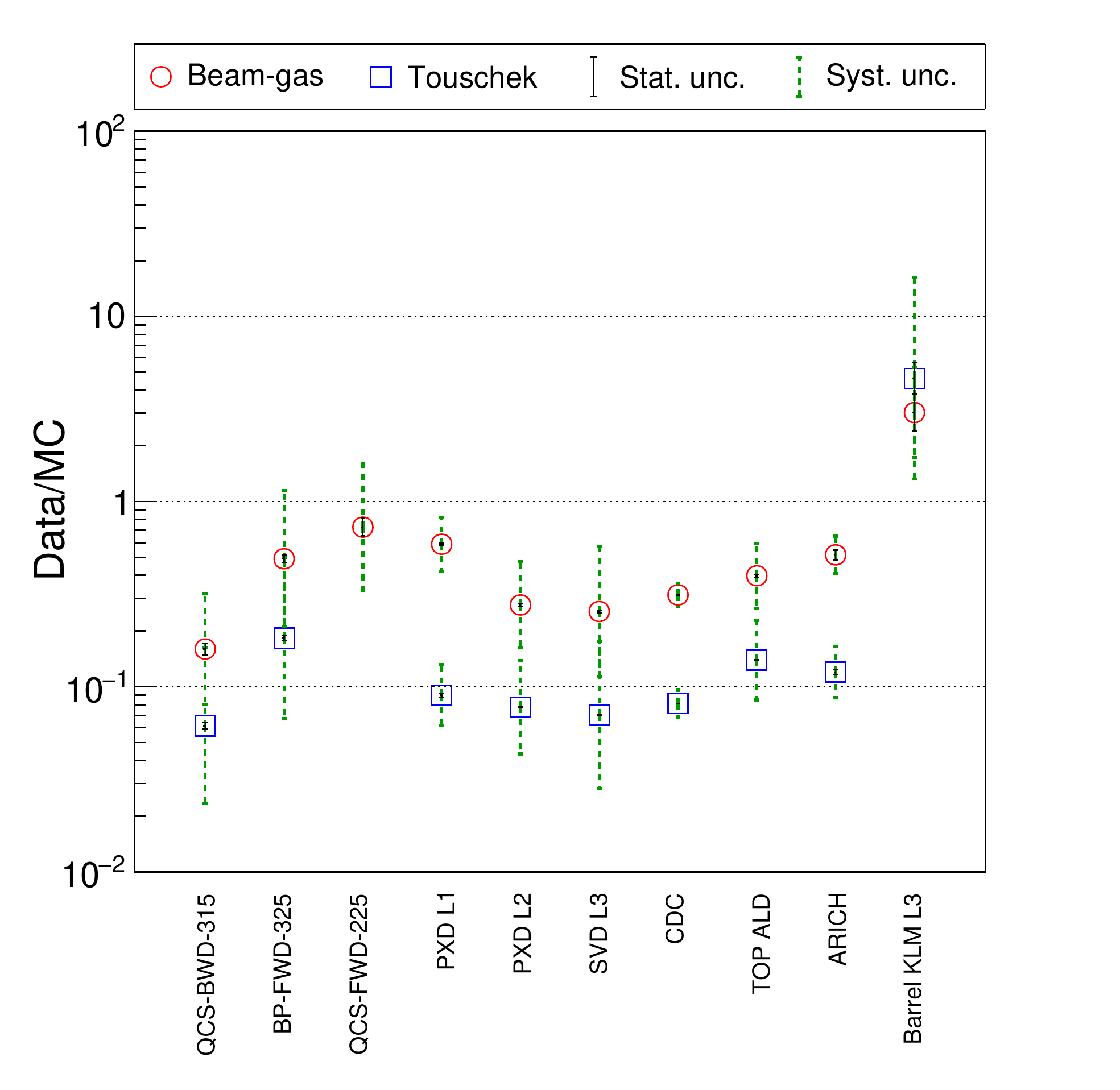}}
\\
\subfloat[\label{secD:fig5c}Luminosity background.]{\includegraphics[width=0.5\linewidth]{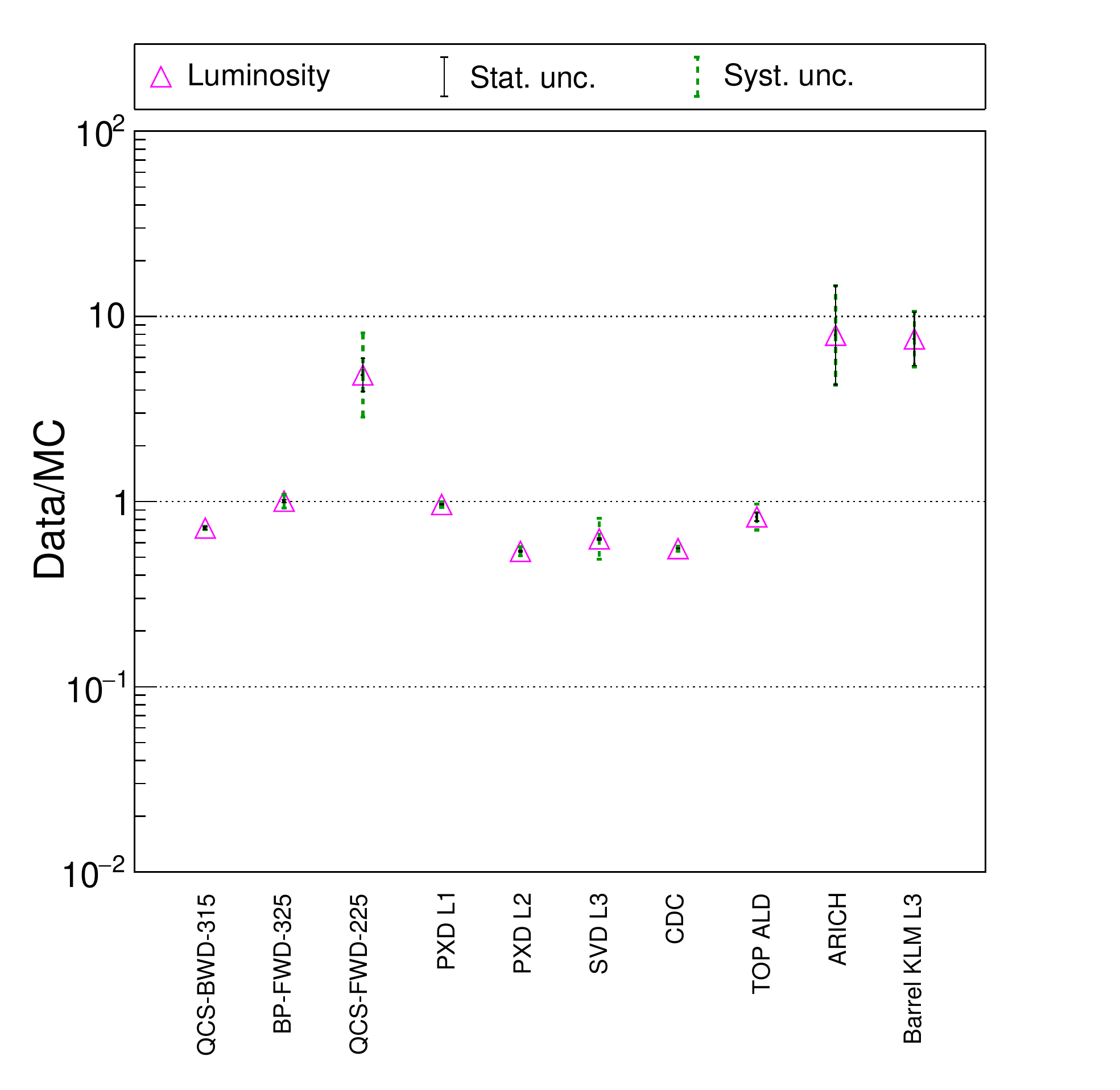}}
\caption{\label{fig:fig5}Combined Data/MC ratios for LER~(a), HER~(b) single-beam and Luminosity~(c) backgrounds.}

\end{figure}
%%%%%%%%%%%%%%%%%%%%%%%%%%%%%%%%%%%%%%%%%%%%%%%%%%%%%%%%%%%%%%%%%%%%%%%%%%%
%%%%%%%%%%%%%%%%%%%%%%%%%%%%%%%%%%%%%%%%%%%%%%%%%%%%%%%%%%%%%%%%%%%%%%%%%%%
\begin{figure}[htbp]
\begin{center}
\includegraphics[width=\linewidth]{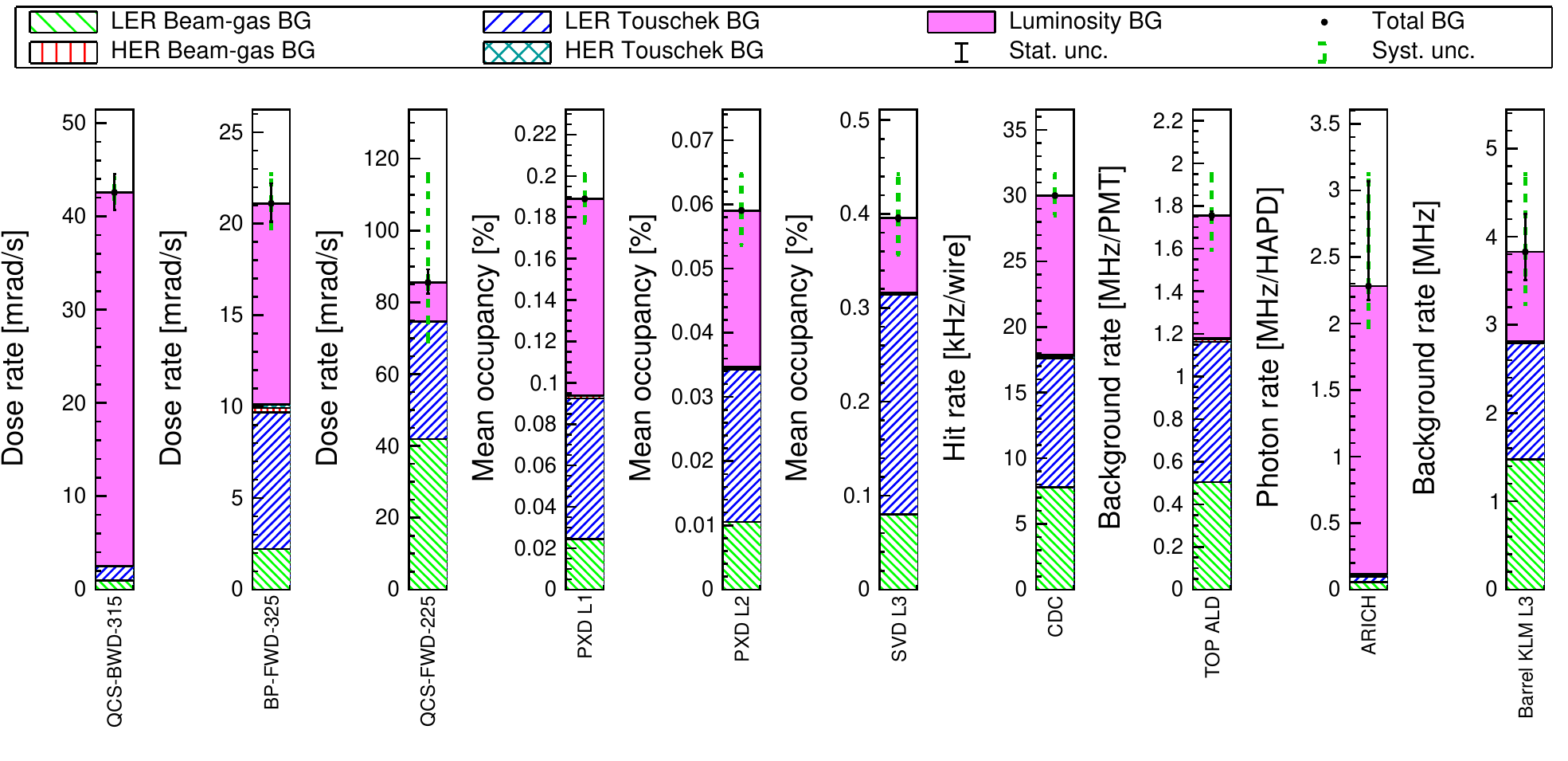}
\end{center}
\caption{\label{fig:fig6}Estimated Belle~II backgrounds for Setup-1. Each column is a stacked histogram.}

\end{figure}
%%%%%%%%%%%%%%%%%%%%%%%%%%%%%%%%%%%%%%%%%%%%%%%%%%%%%%%%%%%%%%%%%%%%%%%%%%%
%%%%%%%%%%%%%%%%%%%%%%%%%%%%%%%%%%%%%%%%%%%%%%%%%%%%%%%%%%%%%%%%%%%%%%%%%%%
\begin{figure}[htbp]
\begin{center}
\includegraphics[width=\linewidth]{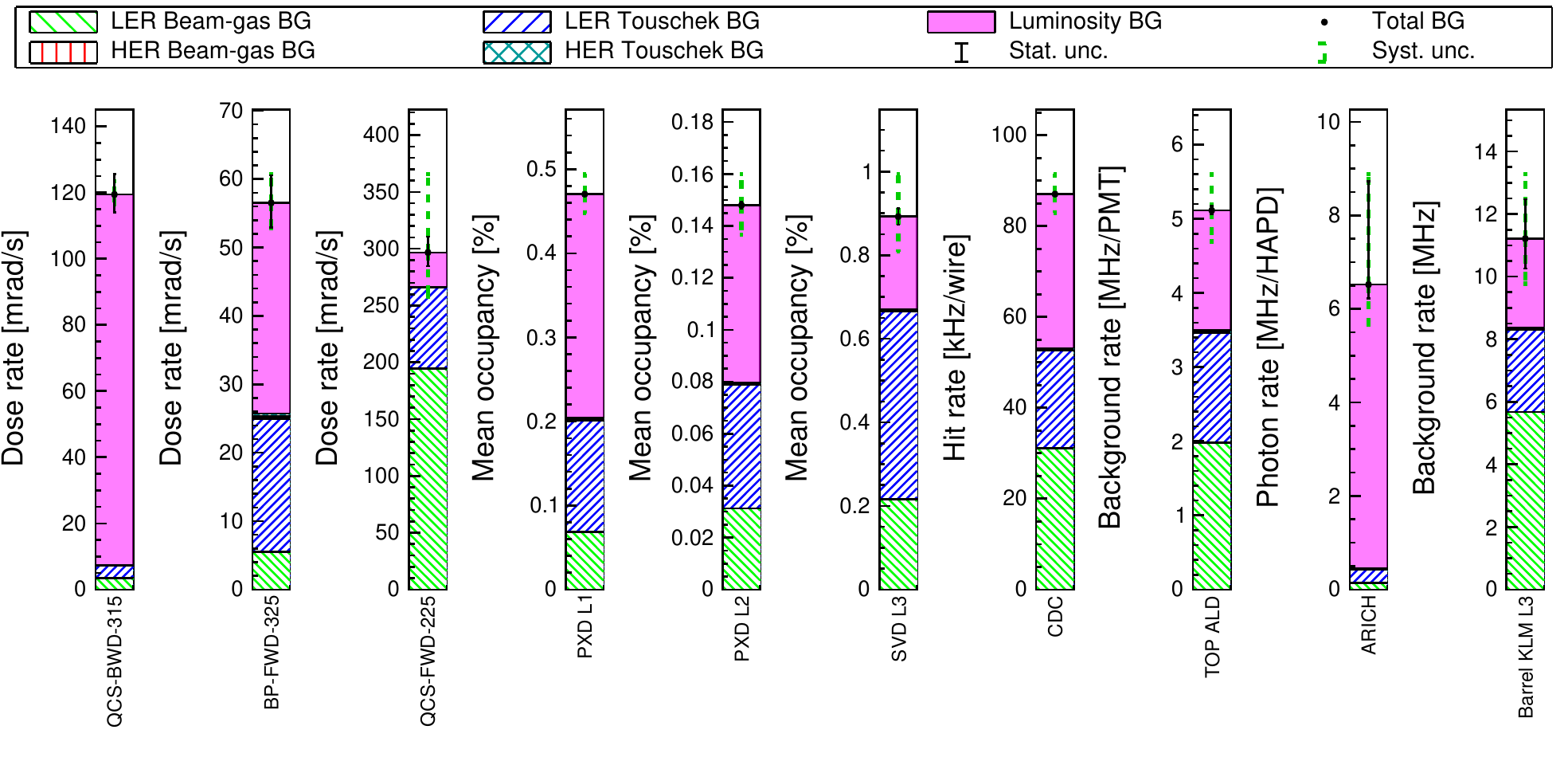}
\end{center}
\caption{\label{fig:fig7}Estimated Belle~II backgrounds for Setup-2. Each column is a stacked histogram.}

\end{figure}
%%%%%%%%%%%%%%%%%%%%%%%%%%%%%%%%%%%%%%%%%%%%%%%%%%%%%%%%%%%%%%%%%%%%%%%%%%%
%%%%%%%%%%%%%%%%%%%%%%%%%%%%%%%%%%%%%%%%%%%%%%%%%%%%%%%%%%%%%%%%%%%%%%%%%%%
\begin{figure}[htbp]
\begin{center}
\includegraphics[width=\linewidth]{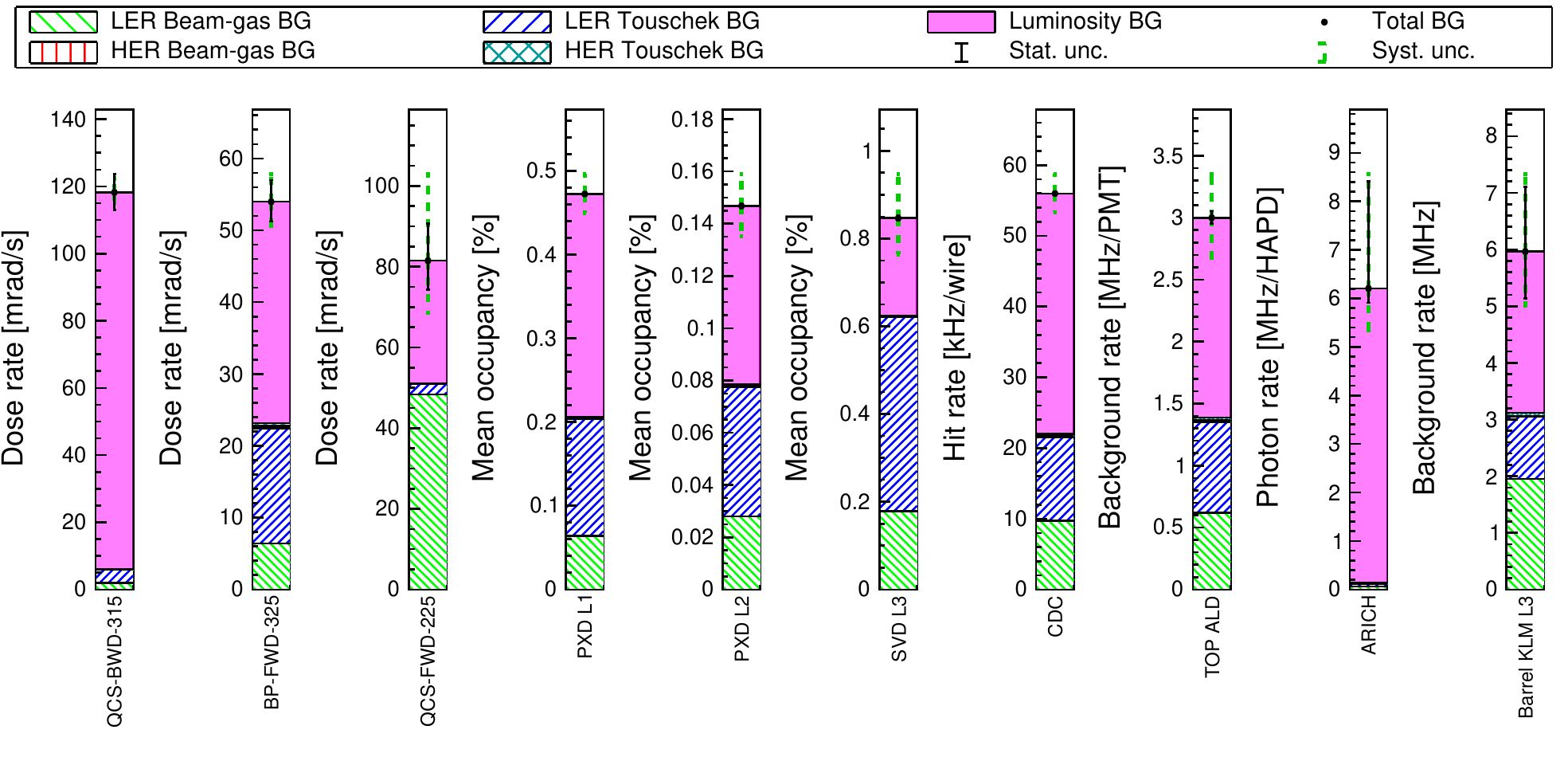}
\end{center}
\caption{\label{fig:fig8}Estimated Belle~II backgrounds for Setup-2* with the NLC. Each column is a stacked histogram.}

\end{figure}
%%%%%%%%%%%%%%%%%%%%%%%%%%%%%%%%%%%%%%%%%%%%%%%%%%%%%%%%%%%%%%%%%%%%%%%%%%%
%%%%%%%%%%%%%%%%%%%%%%%%%%%%%%%%%%%%%%%%%%%%%%%%%%%%%%%%%%%%%%%%%%%%%%%%%%%
%\begin{figure}[htbp]
%\begin{center}
%\includegraphics[width=\linewidth]{BackgroundComposition_betay03.pdf}
%\end{center}
%\caption{\label{fig:fig9}Estimated Belle~II background for Setup-3. Each column is a %stacked histogram.}

%\end{figure}
%%%%%%%%%%%%%%%%%%%%%%%%%%%%%%%%%%%%%%%%%%%%%%%%%%%%%%%%%%%%%%%%%%%%%%%%%%%
%%%%%%%%%%%%%%%%%%%%%%%%%%%%%%%%%%%%%%%%%%%%%%%%%%%%%%%%%%%%%%%%%%%%%%%%%%%
\begin{figure}[htbp]
\begin{center}
\includegraphics[width=\linewidth]{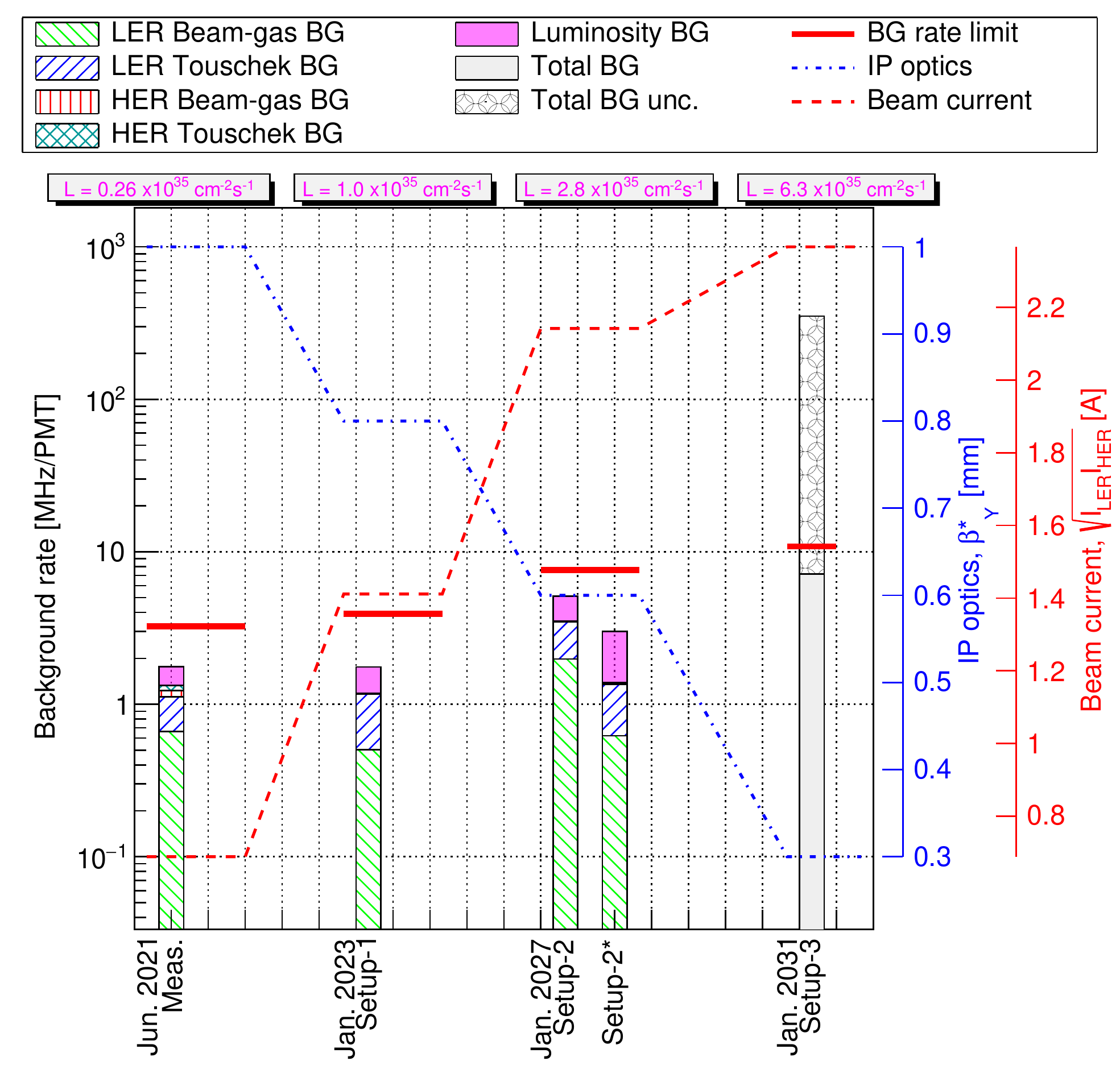}
\end{center}
\caption{\label{fig:fig10}Measured and predicted beam background hit rates in the ALD-type MCP-PMTs of the Belle~II TOP detector, for various machine configurations. These PMTs are expected to be the detector components most vulnerable to beam backgrounds. Predicted backgrounds are acceptable in all scenarios except Setup-3, where there are large systematic uncertainties.}

\end{figure}
%%%%%%%%%%%%%%%%%%%%%%%%%%%%%%%%%%%%%%%%%%%%%%%%%%%%%%%%%%%%%%%%%%%%%%%%%%%
\end{document}